\documentclass[conference]{IEEEtran}
\IEEEoverridecommandlockouts

\usepackage{cite}
\usepackage{amsmath,amssymb,amsfonts, physics}
\usepackage{newtxtext,newtxmath}
\usepackage{algorithmic}
\usepackage{hyperref}
\usepackage{caption}
\usepackage{romannum}
\usepackage{float}
\usepackage{comment}
\usepackage{lipsum}
\usepackage{graphicx}
\usepackage{textcomp}

\usepackage[version=4]{mhchem}
\usepackage{xcolor} 
\def\BibTeX{{\rm B\kern-.05em{\sc i\kern-.025em b}\kern-.08em
		T\kern-.1667em\lower.7ex\hbox{E}\kern-.125emX}}
\usepackage{fancyhdr}
\begin{document}
	
	\title{Multifunctional chalcogenide (\ce{As2S3}, \ce{As2Se3}) dual-core PCF design for mid-IR optical communications}

	\author{\IEEEauthorblockN{A K M Sharoar Jahan Choyon\textsuperscript{*,1}, Ruhin Chowdhury\textsuperscript{*}}
		\IEEEauthorblockA{\textsuperscript{*}\textit{Dept. of Physics \& Astronomy, The University of New Mexico, Albuquerque, NM 87131, USA} \\
		\textsuperscript{1}{choyon@unm.edu, choyonsharoar@gmail.com}  
		}}

	\maketitle

	\begin{abstract}
	This work presents an integrated design of a multifunctional hexagonal lattice dual-core photonic crystal fiber (DC-PCF) with elliptical air-hole surrounding the cores, using two different chalcogenides (\ce{As2S3}-Arsenic sulfide and \ce{As2Se3}-Arsenic selenide) for mid-infrared (mid-IR) optical communications. Numerical study of both chalcogenide DC-PCF structures exhibits that the DC-PCFs are highly birefringent and single-moded in the mid-IR wavelength region (5 $\mu m$ to 13 $\mu m$) of optical communications. Results show that both DC-PCFs can be operated as polarization splitters at 9 $\mu m$ wavelength for two orthogonal modes and they can also be used as WDM MUX-DeMUX for both X- and Y-polarization fundamental supermodes separating 9 $\mu m$/10 $\mu m$ wavelengths (\ce{As2S3} DC-PCF) and 8 $\mu m$/9 $\mu m$ wavelengths (\ce{As2Se3} DC-PCF). Moreover, the proposed chalcogenide DC-PCFs, with much lower splice loss at 8-10 $\mu m$ wavelength region, are appropriate enough for the practical applications in integrated optics and photonics.

	\end{abstract}


\section{Introduction}

Photonic crystal fibers (PCFs) have been the topic of extensive research in recent years because of their wide range of applications, flexibility of the design, and distinctive advantages in controlling light as well as numerous exclusive features, such as their large effective mode area, high nonlinearity, adjustable dispersion, large birefringence, and constant single-mode nature, as well as maintaining polarization at mid-IR wavelengths \cite{1}-\cite{4}. Many optical devices such as polarization splitter/filter, coupler, and wavelength coupler/splitter can be accomplished using this PCF, which performs an important role in integrated optical systems \cite{5}-\cite{7}. In the polarization splitters, optical signal is normally separated into two orthogonal polarization components \cite{choyon1}-\cite{choyon8}. Aside from these applications, it can also be used in high-speed radio-over-free-space optical technology \cite{7a}. Therefore, multifunctional fiber has gained major concentration by the scientists and engineering community for the application of integrated and compact optical systems. In the area of optical communication, many photonic and optical devices such as wavelength division MUX-DeMUX, polarization splitter have been thoroughly researched that result in the invention of newly designed photonic crystal fiber for their unique tunable characteristics \cite{8}. 

Besides, many research investigations have been done to depict multifunctional DC-PCF \cite{8}-\cite{9}, that contains maintained desirable properties for optical communications, for example, design flexibility, short coupling length compared to traditional fiber optic coupler \cite{5}, large birefringence value, low splice loss, endlessly operation in single-mode \cite{10}-\cite{11}, effective polarization, and also wavelength splitting \cite{12}. A PCF multi-core fiber can additionally provide short coupling lengths in the millimeter range. A PCF with these properties is an attractive candidate for both MUXing (mixing the different wavelengths incoming) and DeMUXing (dividing the power into multiple wavelengths) \cite{4}.

Currently, mid-IR photonic devices have been urged significant interest due to its potential applications in free space optical communication, spectroscopic sensing, imaging and so on \cite{13}-\cite{14}. Mid-IR optical devices, in particular, are widely used for their various capabilities, including power coupling, wavelength splitting, and multiplexing \cite{15}. A variety of approaches have now been utilized to develop these devices, including the use of multi-mode interference (MMI) waveguide coupling, glass laser inscription waveguides, and mid-IR fibers \cite{16}-\cite{17}. In particular, due to the consequence of their practically achievable Fermi level, MMI waveguide couplers are limited in their operating bandwidth. As a means to mitigate this issue, mid-IR photonic crystal fibers (PCFs) have been found to be the best choice for developing couplers and multiplexers due to their short coupling length and greater design flexibility \cite{17}-\cite{18}.

Due to their flexible structural design, some PCF-based polarization splitters have improved performance in the near-infrared region (0.7-2 $\mu m$). Fan et al. \cite{19}, in 2015, improved the coupling properties of a soft glass DC-PCF with a fiber length of 52.29 $mm$ by using a high refractive index \ce{As2S3} core. It has been reported that Zhao et al. \cite{20}, in 2016, designed a polarization splitter with two fluorine-doped cores and achieved a fiber length of 52.8 $mm$. According to Hagras et al. (2019) \cite{10}, chalcogenide glass (\ce{As2S3}) and nematic liquid crystal were used to achieve 1.55 $\mu m$ and 1.30 $\mu m$ polarization splitters with fiber lengths of 83 $\mu m$ and 166 $\mu m$, respectively. According to a recent study by Rahman et al. \cite{8}, in 2019, an elliptical DC-PCF was used along with silica material to achieve a polarization splitter at 1550 $nm$ with a fiber length of 39.8 $mm$, and it was used in WDM MUX-DeMUX for separating 1300 $nm$ and 1550 $nm$ wavelengths in X and Y polarization modes at 9.9 $mm$ and 16.65 $mm$ fiber lengths, respectively. All of the above research (ref. \cite{8}, \cite{10}, \cite{19}-\cite{20}) were conducted in the near-infrared region (0.7-2 $\mu m$), whereas, this proposed work, for the first time, has been conducted in the mid-IR region (5-13 $\mu m$) using chalcogenide elliptical DC-PCFs with an improved shorter coupling length, shorter fiber length for polarization splitter and WDM MUX-DeMUX, low splice loss for the practical applications in integrated and compact optical and photonic communications systems as a short-length multifunctional device.

The remaining sections of the paper are arranged as follows. At the beginning of Section \ref{sys}, it elucidates the explanations of choosing \ce{As2S3} and \ce{As2Se3}  as the two different background materials for our proposed DC-PCF design. Section \ref{design} elaborates the design parameters along with the simulation methodology using COMSOL Multiphysics and explains the linear optical properties of DC-PCF. Section \ref{appl} shows the possible applications of our proposed DC-PCF to use it as a short-length multifunctional device and finally, the paper is summarized in Section \ref{conclusion}. 

\section{\textbf{Material Selection for mid-IR}}\label{sys}
    PCFs in the mid-IR range have been designed so far with transparent fluoride, telluride, and chalcogenide glasses (ChG). In particular, ChGs exhibit low losses and a high level of transparency up to a wavelength of 25 $\mu m$ while possessing a nonlinearity of 1000 times that of silica. As a result, they can be used as infrared optical fibers or waveguides \cite{23}-\cite{24}. Aside from the wide transmission windows in the mid-IR range, ChGs also have very large nonlinear and linear refractive indices \cite{25}.
	
	In our study, we have used ChGs as background material for the PCF. The chalcogens in ChGs include one or more elements from the periodic table (e.g., sulphur, selenium, tellurium, except for oxygen) that are covalently bonded to other elements such as As, Ge, Sb, Ga, Si, or P. In this study, we have used \ce{As2Se3} and \ce{As2S3} as chalcogenide-based background materials. The reasons for using these chalcogenides as background materials are that they have a wider transmission window than silica, a relatively constant material loss, and they are more practical in use \cite{26}. Infrared transmission, optical fibers for optical sensors, and telecommunication applications benefit from the low intrinsic material loss of \ce{As2Se3} chalcogenide glass \cite{27}. With an attenuation coefficient of less than 1 $cm^{-1}$, \ce{As2Se3} glass has excellent optical transparency between 0.85 and 17.5 $\mu m$ \cite{28}. The properties of ChGs based on \ce{As2S3} have attracted intense research attention due to their high Kerr nonlinearity $n_2$ (100-500 times greater than silica glass), low linear loss, and small two-photon absorption (TPA) coefficient \cite{29}. In addition, here we have used the specifications of the commercially available Femtofiber Dichro mid-IR laser source from \textit{Toptica Photonics} to generate light with a tunable wavelength in the range of 5-13 $\mu m$ (23-60 THz) with a repetition rate of 80 MHz and an emission bandwidth greater than 400  $cm^{-1}$ \cite{30}.

\section{\textbf{Design and Analysis of DC-PCF}}	\label{design}
		
\begin{figure}
	\centering
	\includegraphics[width=3.5in,height=3in,keepaspectratio]{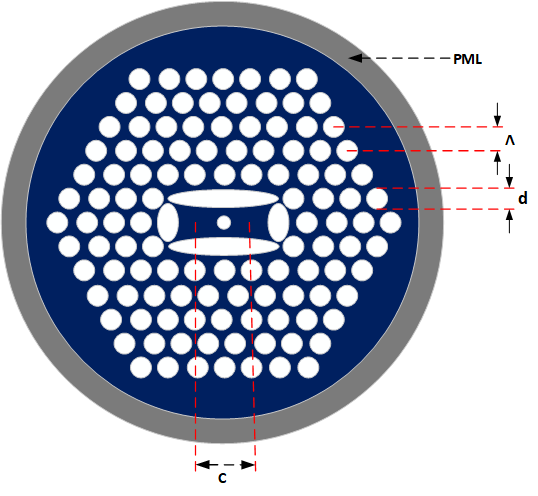} 
	\caption{ A 2D cross-section of the proposed chalcogenide DC-PCF incorporating elliptical air holes surrounding the cores with PML at the boundary, and $C=2\Lambda$. In the big elliptical air-hole: Major (horizontal) axis= 14 $\mu m$, Minor (vertical) axis= 2.8 $\mu m$; in the small elliptical air-hole: Major (vertical) axis= 5.4 $\mu$m, Minor (horizontal) axis= 2.8 $\mu m$. In addition to elliptical air holes, the structure has two more types of air holes, inside (center) and outside of elliptical air-holes. The diameter of- outer air-hole, $d= 2.8$ $\mu m$, and center air hole= 1.9 $\mu m$. Here, the separation between outer air holes (center to center distance or pitch), $\Lambda$=3.5 $\mu m$ and, the DC-PCF cores are separated by $C$. }\label{PCF}	
    \end{figure}

\begin{figure*}
	\centering
	\includegraphics[width=6in,height=5in,keepaspectratio]{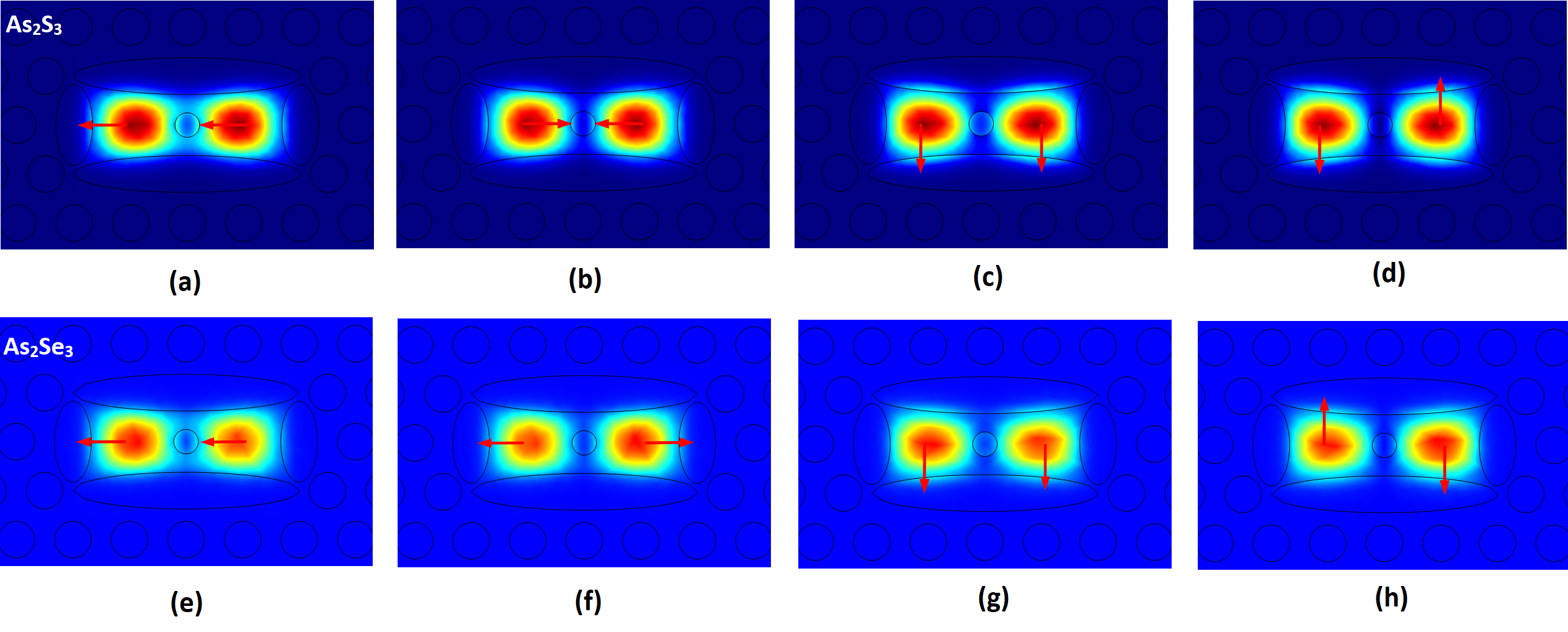} 
	\caption{\centering Electric field distribution of the fundamental supermodes for the proposed \ce{As2S3} and \ce{As2Se3} DC-PCFs of FIG. \ref{PCF} simulated at 8 $\mu m$. (a) X-even (\ce{As2S3}), (b) X-odd  (\ce{As2S3}), (c) Y-even  (\ce{As2S3}), (d) Y-odd  (\ce{As2S3}), (e) X-even (\ce{As2Se3}), (f) X-odd  (\ce{As2Se3}), (g) Y-even  (\ce{As2Se3}), and (h) Y-odd  (\ce{As2Se3}). }\label{pol}	
    \end{figure*}

In Fig. \ref{PCF}, our proposed structure for DC-PCF is shown which has holes of elliptical shape surrounding the cores. The design has outer air hole with diameter, $d = 2.8$ $\mu m$; pitch, $\Lambda$ = 3.5 $\mu m$ and core separation, $C = 2\Lambda$ = 7 $\mu m$. As shown in the figure, the major and minor axes are 14 $\mu m$ and 2.8 $\mu m$ for the elliptical air-hole parallel to the axis connecting the cores, and they are 5.4 $\mu m$ and 2.8 $\mu m$ for the elliptical air-hole perpendicular to the axis connecting the cores, respectively. The process “stack-and draw” is used for manufacturing of such elliptical air hole PCF as presented in Ref. \cite{31}. To simulate the design numerically, COMSOL Multiphysics has been used, whereas the finite element method (FEM) was applied to all triangular edge elements based on cylinder perfectly matched layer (PML) boundary conditions. The structure of FIG PCF has been developed on the basis of the structures in \cite{8} and \cite{32}. We demonstrate light propagation through DC-PCFs through four different types of supermodes, which differ with respect to electric field distribution with regard to horizontal and vertical polarization. Fig. \ref{pol} illustrates the fundamental polarization modes for \ce{As2S3} and \ce{As2Se3} DC-PCF.

	 \subsection{\textbf{Effective Refractive Index}}  

     To incorporate the result of both waveguide and material dispersion, refractive index of chalcogenide materials (both \ce{As2S3} and \ce{As2Se3}) is calculated from the Sellmeier formula \cite{35}:

    \begin{equation} \label{neff}
        \epsilon_{r}(\lambda) = 1+ \sum_{n=1}^{3}\frac{A_{n}\lambda^{2}}{\lambda^{2} - B_{n}}
    \end{equation} 
    Here, the value of wavelength ($\lambda$) is noted in $\mu m$ and $A_{n}$ and $B_{n}$ are called Sellmeier coefficients and the corresponding value of $A_{n}$ and $B_{n}$ for $n = 1, 2, 3$ is given in Table \ref{tab1}.
    
    \begin{table*}
	\centering
	\caption{Constant $A_{n}$ and $B_{n}$ ($\mu$$m^{2}$) ($n = \text{1, 2 and 3}$) for \ce{As2S3}  and \ce{As2Se3}, used in Thompson’s Sellmeier Equation \cite{26a}, \cite{31a}.}\label{tab1}
\begin{tabular}{c c c c c c c}
			\hline
			\textbf{Material} & \textbf{$A_{1}$} & \textbf{$A_{2}$} & \textbf{$A_{3}$} & \textbf{$B_{1}$($\mu$$m^{2}$)} & \textbf{$B_{2}$($\mu$$m^{2}$)} & \textbf{$B_{3}$($\mu$$m^{2}$)}\\ \hline
		 
			\ce{As2S3} & 1.8983 & 1.9222 & 0.8765 & 0.0225 & 0.0625 & 0.1225\\
			\ce{As2Se3}  & 4.994 & 0.120 & 1.710 & 0.0583 & 361 & 0.2332\\

			\hline
			
		\end{tabular}
	
\end{table*}

	\begin{figure*}
	\centering
	\includegraphics[width=7in,height=7in,keepaspectratio]{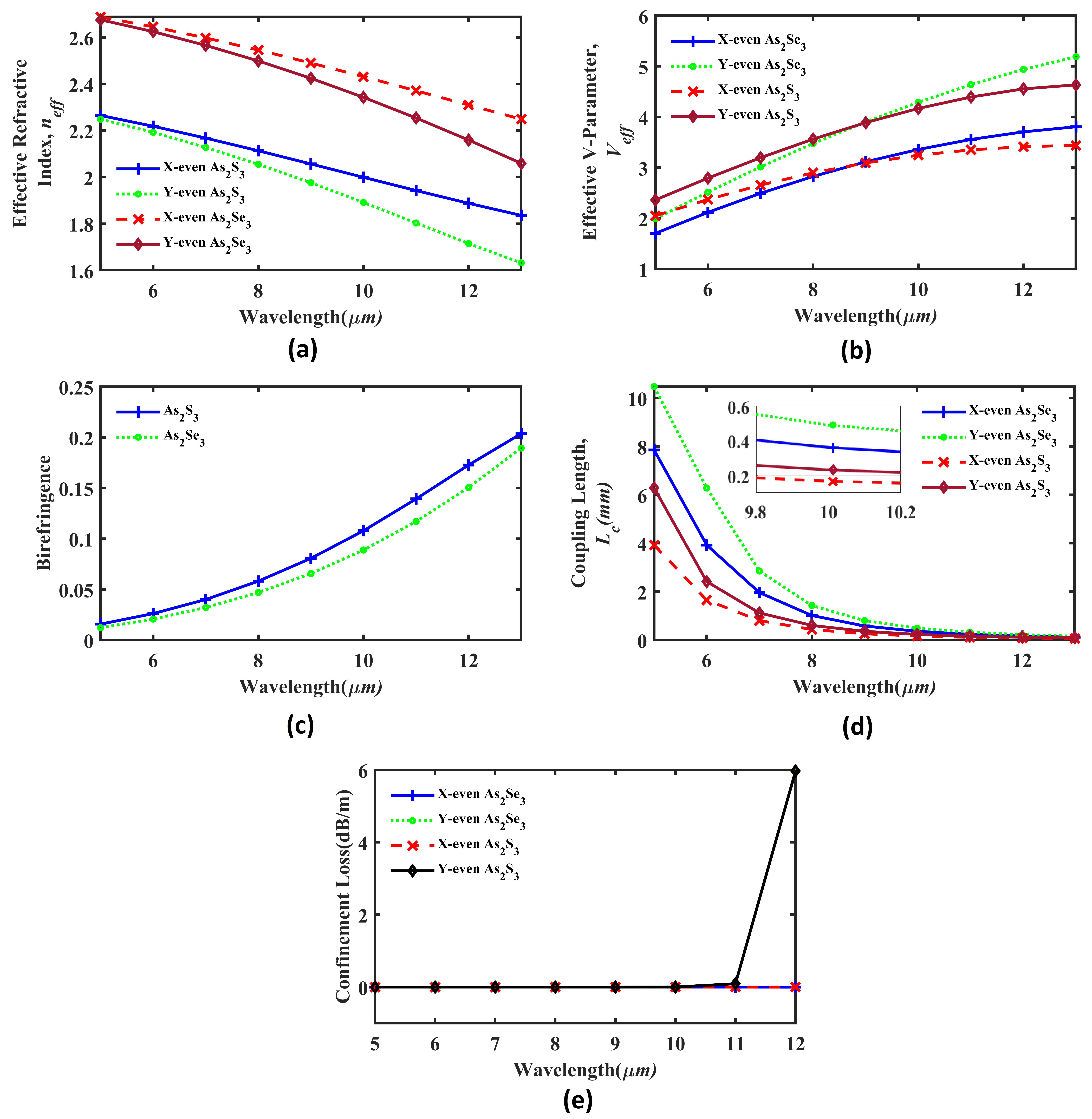} 
	\caption{\centering Linear optical properties of proposed DC-PCF for both chalcogenide (\ce{As2S3} and \ce{As2Se3}) materials: (a) Effective refractive index (b) Effective V-parameter (c) Birefringence properties (d) Coupling length (e) Confinement loss. }\label{nvbl}	
    \end{figure*}

    In Fig. \ref{nvbl}(a), the effective refractive index ($n_{\text{eff}}$) of even symmetry for both horizontal and vertical polarizations of both chalcogenide materials have been plotted with respect to operating wavelength. According to the figure, due to a larger diffused electric field distribution, light confinement within the core regions increases with wavelength, reducing the $n_{\text{eff}}$. The Fig. \ref{nvbl}(a) additionally indicates that as wavelength increases, the difference between the effective indices of X and Y supermodes increases, resulting in increased birefringence in the larger wavelength region.

 \subsection{\textbf{Effective V-Parameter}}
   
   The Normalized Frequency Parameter of a fiber, also called V-parameter, is an important parameter for characterising PCF. The value of effective V-parameter, $V_{\text{eff}}$ can be computed from the following equation \cite{11}:

	\begin{equation}
        V_{\text{eff}} = 2\pi\frac{\Lambda}{\lambda}\sqrt{n_{0}^{2} - n_{\text{eff}}^{2}}
    \end{equation}
    
According to Sellmeier Equation (\ref{neff}), $n_{0}$ is the chalcogenide's refractive index, while $n_{\text{eff}}$ is its effective refractive index calculated using modal analysis. In Fig. \ref{nvbl}(b), we can clearly notice the change of effective V parameter with the change of wavelength ($\lambda$). The requirement for single mode operation in a particular hexagonal or triangular lattice PCF is $V_{\text{eff}}\leq4.1$ \cite{11}. Thus, Fig. \ref{nvbl}(b) illustrates that the DC-PCF we propose is single-moded at 5-13 $\mu m$ wavelengths for \ce{As2S3}, and single-moded for the region of wavelength upto 10 $\mu m$ for \ce{As2Se3}, although \ce{As2Se3} deviates from the single-mode nature after 10 $\mu m$ according to the ref. \cite{11}. Moreover, the requirement of $V_{\text{eff}}$ for single mode operation depends on the geometrical shape of PCF, pitch, diameter to pitch ratio, wavelength, etc., where our proposed structure for dual-core PCF in mid-IR region is completely different from ref. \cite{11}. Additionally, the confinement loss of both \ce{As2S3} and \ce{As2Se3} is zero for the region of wavelength upto 10 $\mu m$, depicted in Fig. \ref{nvbl}(e), where the  attenuation or confinement loss due to optical field leakage to the claddings of waveguide structures that propagate light is calculated using the following formula \cite{36}:

\begin{equation}
        \text{Confinement loss} = \frac{40\pi}{\ln{(10)}\lambda}\text{Im}(n_{\text{eff}})   \hspace{2mm}\text{(dB/m)}
    \end{equation}

Here, the imaginary component of the effective refractive index is Im($n_{\text{eff}}$). Hence, from the Fig. \ref{nvbl}(b), we can confirm that our proposed DC-PCFs are single moded over the wavelength region considered upto 10 $\mu m$ and can be used for wavelength windows for optical communications.  
   
   \subsection{\textbf{Coupling Length}}
   The length of fiber at where optical power fully lifted from one core to another is represented as coupling length ($L_c$) which is derived from the following equation \cite{32}:
   
   \begin{equation}
        L_{C} = \frac{\pi}{\beta_{even} - \beta_{odd}}
    \end{equation}
    Where, $\beta_{even}$ and $\beta_{odd}$ are the symmetric and anti-symmetric orthogonally polarized supermode propagation constants, respectively.
    According to Fig. \ref{nvbl}(d), the coupling length decreases as the wavelength of the fiber increases. The coupling lengths of horizontal $X_{even}$/vertical $Y_{even}$ supermodes for \ce{As2S3} and \ce{As2Se3} are 0.249 mm/0.360 mm, and 0.992 mm/1.496 mm, respectively at 9 $\mu m$ wavelength. Fig. \ref{pol}(a) and Fig. \ref{pol}(e) illustrate the phenomenon of the short coupling length of the X-supermode. This is because the two cores are strongly coupled via horizontal channels in the core regions. Due to the fact that both cores are aligned parallel to the x-axis, X-polarized mode has a shorter coupling length than Y-polarized mode. Additionally, X-polarized mode has higher coupling than Y-polarized mode. The coupling  characteristic of DC-PCF has a  potential application  in  the  wavelength  selective system, i.e. a device can be used as a MUX-DEMUX or power-coupler in the WDM system with short coupling length compared to regular fiber. Moreover, it is required to decrease the coupling length for compensating birefringence with low confinement loss to demonstrate polarization-insensitive properties and for the use in wide-band optical communications \cite{2},\cite{4},\cite{32}.

    \subsection{\textbf{Coupling Length Ratio and Birefringence}}
\begin{figure}
	\centering
	\includegraphics[width=3.4in,height=7.5in,keepaspectratio]{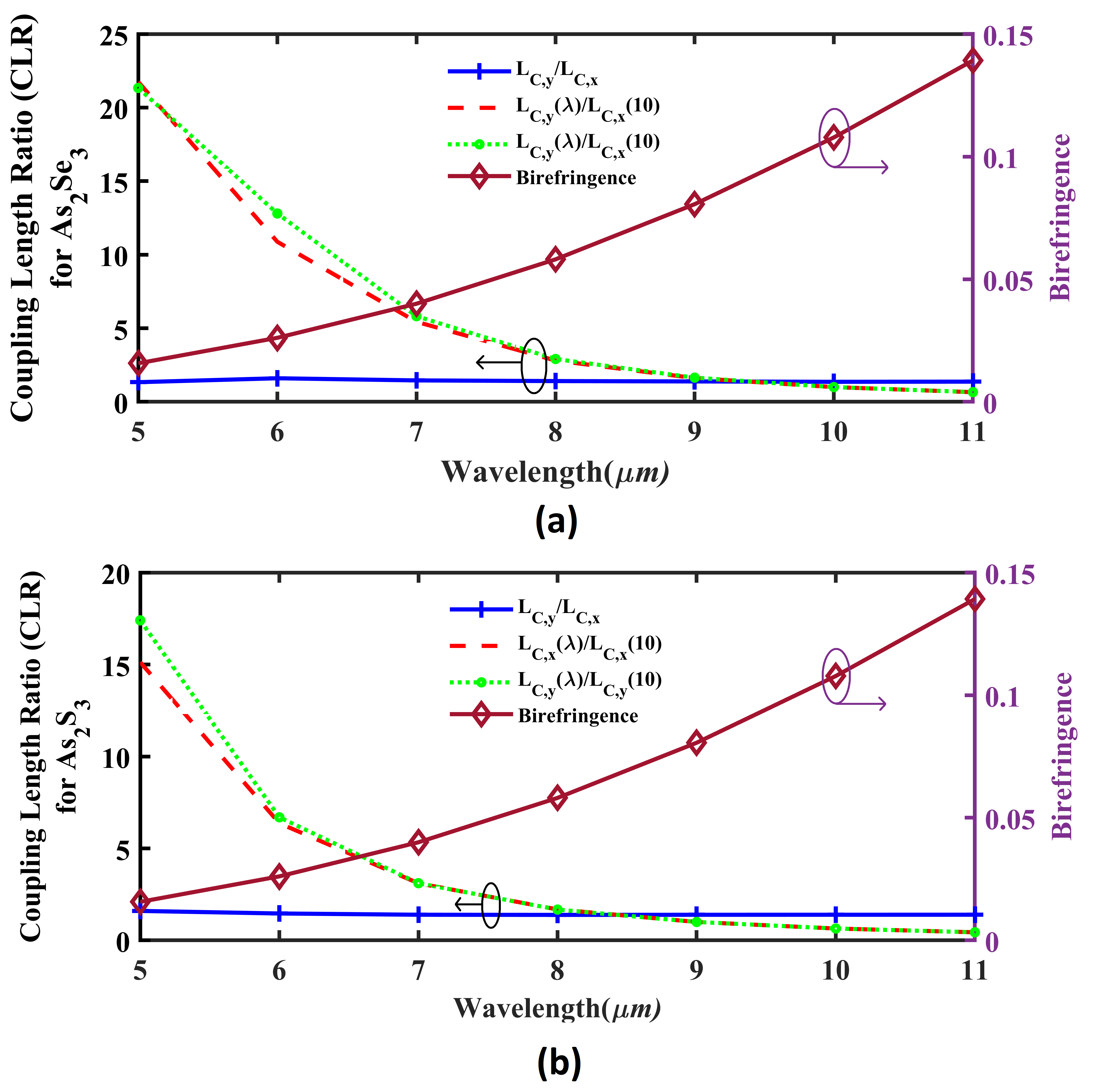} 
	\caption{Coupling Length Ratio (CLR), $L_{C,\lambda}/L_{C,10}$ and birefringence of the proposed DC-PCFs for: (a) \ce{As2Se3} (b) \ce{As2S3}  chalcogenide materials. }\label{clr}	
    \end{figure}

    Coupling length ratio (CLR) is one key parameter to calculate the felicity of a coupling structure as a polarization splitter. This ratio can be expressed as follows \cite{1}:
    \begin{equation}
        CLR = \frac{L_{C,y}}{L_{C,x}}
    \end{equation}
    Here, the coupling length for X-mode and Y-mode is $L_{C,x}$ and $L_{C,y}$, respectively. The two orthogonal modes in a fiber cannot be separated unless the fiber lengths for the two modes satisfy the following condition: $L = mL_{C,x} = nL_{C,y}$, where $m$ and $n$ are positive integers and their parity varies (i.e., even or odd) within each mode \cite{33}. CLR values are depicted for the X and Y supermodes of the proposed DC-PCFs, as illustrated in Fig. \ref{clr}. At 8 $\mu m$ and 9 $\mu m$, the CLR for \ce{As2Se3} is found 1.454 (or 16/11) and 1.50 (or 15/10), respectively. Similarly, the CLR for \ce{As2S3} is found 1.441 (or 36/25) and 1.437 (or 23/16), respectively at 9 $\mu m$ and 10 $\mu m$.

   Using a PCF coupler, two wavelengths $\lambda_1$ and $\lambda_2$ can be effectively separated provided that the coupling lengths $L_{C,\lambda_1}$ and $L_{C,\lambda_2}$ at wavelengths $\lambda_1$ and $\lambda_2$, respectively, fulfil the following conditions: \cite{32}:

    \begin{equation}
        \frac{L_{C,\lambda_1}}{L_{C,\lambda_2}} = \frac{\text{even integer}}{\text{odd integer}}
    \end{equation}
    
    or,
    \begin{equation}
        \frac{L_{C,\lambda_1}}{L_{C,\lambda_2}} = \frac{\text{odd integer}}{\text{even integer}}
    \end{equation}
    
     Fig. \ref{clr} shows wavelength-dependent CLRs $L_{C,\lambda_1}/L_{C,\lambda_2}$ for both the X and Y supermodes of DC-PCF structures (\ce{As2Se3} and \ce{As2S3}). The wavelength $\lambda_2$ here in Fig. \ref{clr} is fixed at 10 $\mu$m. In order to get a short-length and high-performance WDM MUX-DeMUX, the optimum value of the ratio  $L_{C,\lambda_1}/L_{C,\lambda_2}$ is to be 1/2 or 2 \cite{1}. From Fig. \ref{clr}, we have found this ratio approximately at 8.5 $\mu m$ for both DC-PCF structures. 
    
    The fiber obtains birefringence, also called double refraction, at the time of different propagation constants $\beta_x$ and $\beta_y$. The property of birefringence occurs when one light beam is split into two isolated beams, with each beam being orthogonally polarized and refracted at a different angle \cite{8}. Fiber birefringence is defined as the difference between the effective refractive indexes of these two polarization states driven by the different propagation phase velocities as shown in the following equation \cite{34}. Moreover, high birefringence is necessary for maintaining two linear orthogonal polarisation states over long distance \cite{34}.
    
      \begin{equation}
        B_m = \frac{| \beta_x - \beta_y |}{2\pi/\lambda} = |n^{x}_{\text{eff}} - n^{y}_{\text{eff}}|
     \end{equation}
     
     In Fig. \ref{clr}, the acquired birefringence of our structures for both the materials (\ce{As2Se3} and \ce{As2S3}) is depicted, which is quite high enough for mid-IR optical communications. 
     
     \section{\textbf{Applications}}\label{appl}
     
     According to the observed data presented above, our proposed DC-PCFs are suited for using as WDM MUX-DeMUX for separating the wavelengths of 8 $\mu m$ and 9 $\mu m$ for \ce{As2Se3} and 9 $\mu m$ and 10 $\mu m$ for \ce{As2S3}. While the CLRs are not ideal for polarization splitting functions, the developed structures for both chalcogenides fulfill the polarization separation criteria and can thus be utilized as polarization splitters.

     \subsection{\textbf{Polarization Splitter}}
     $P_{out,A}$ and $P_{out,B}$ represent the normalized output power generated by cores $A$ and $B$, respectively, when a fundamental mode is launched into core $A$ via the power $P_{in}$ and are as follows \cite{35}:

     \begin{equation}\label{eq8}
        P_{out,A} = P^i_{in}{cos}^2\frac{\pi z}{2L^i_C} 
     \end{equation}
     
     \begin{equation}\label{eq9}
        P_{out,B} = P^i_{in}{sin}^2\frac{\pi z}{2L^i_C} 
     \end{equation}
     Here, $z$= propagation length, $i=x,y$, and $L^i_C$ = coupling length of the $i$-polarized mode.
 
	\begin{figure}
	\centering
	\includegraphics[width=3.4in,height=7.5in,keepaspectratio]{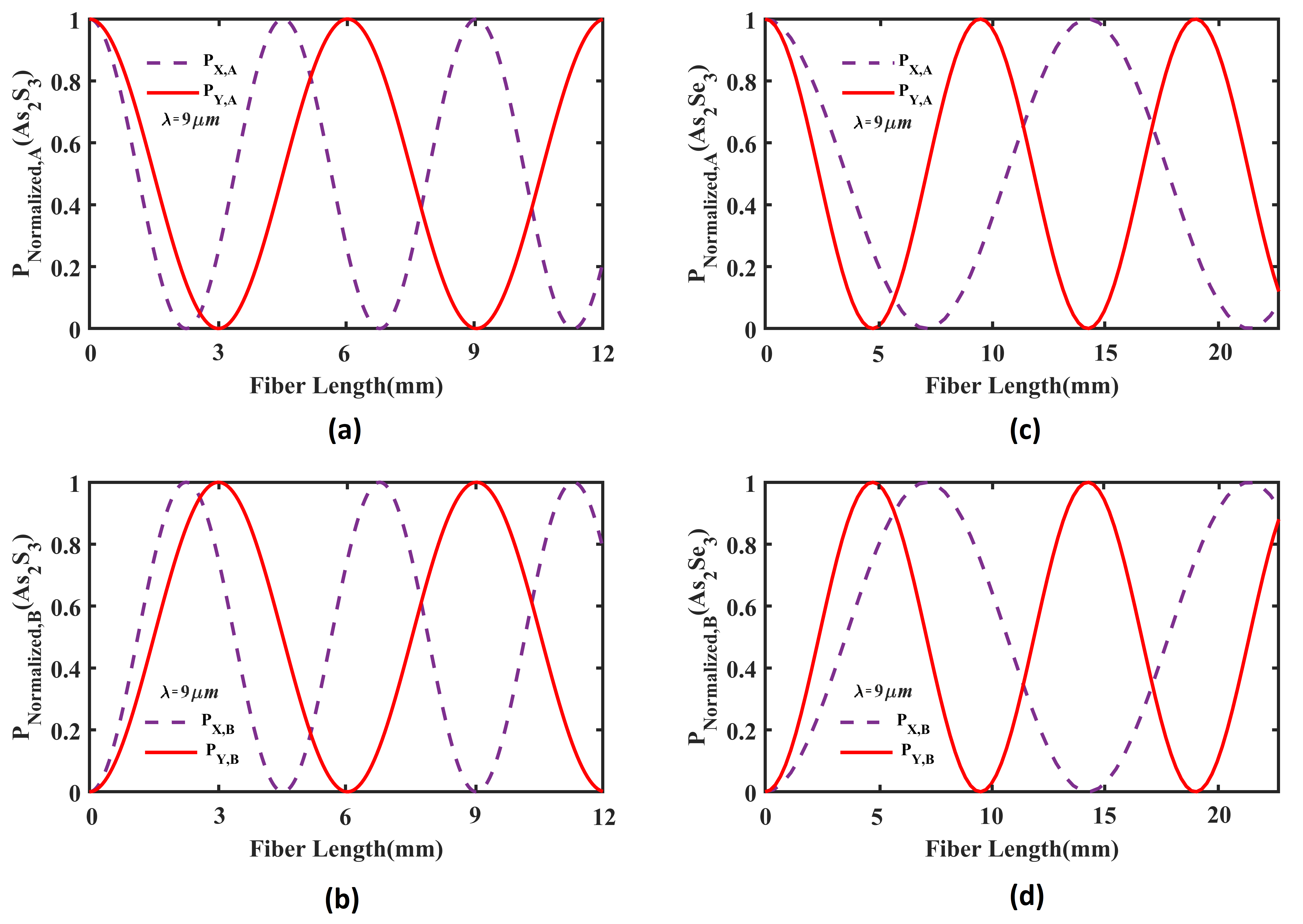} 
	\caption{Normalized output power of DC-PCF structures as a function of propagation length for: (a) core A, \ce{As2S3} (X-mode) (b) core B, \ce{As2S3} (Y-mode) (c) core A, \ce{As2Se3} (X-mode) (d) core B, \ce{As2Se3} (Y-mode). }\label{PBS}	
    \end{figure}
    
    In Fig. \ref{PBS}, the normalized power for different propagation length of core $A$ and core $B$ of both DC-PCF structures mentioned in Fig. \ref{PCF} at 9 $\mu m$ wavelength are shown. From this figure, we have estimated that the DC-PCF structures separate the two orthogonal modes at a distance of $(36 \times 0.249 + 25 \times 0.360)/2 = 8.9$ $mm$ for \ce{As2S3} and at a distance of $(15 \times 0.992 + 10 \times 1.496)/2 = 14.8$ $mm$ for \ce{As2Se3}.
    
	\begin{figure}
	\centering
	\includegraphics[width=3.4in,height=7.5in,keepaspectratio]{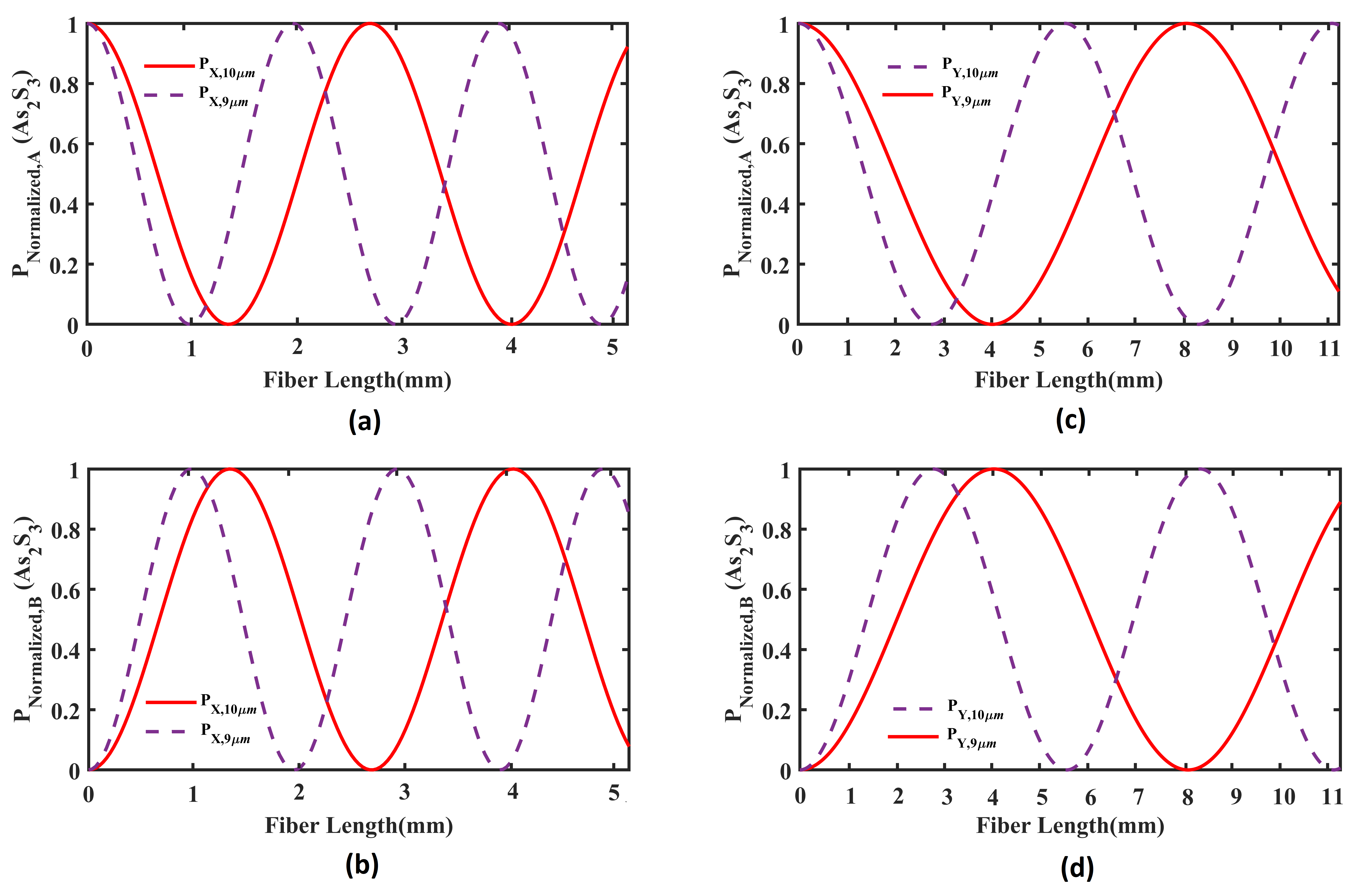} 
	\caption{Normalized output power of \ce{As2S3} DC-PCF structure as a function of propagation length for: (a) X-mode, core $A$ ($\lambda = 9 \mu m$) (b) X-mode, core $B$ ($\lambda = 10 \mu m$) (c) Y-mode, core $A$ ($\lambda = 9 \mu m$) (d) Y-mode, core $B$ ($\lambda = 10 \mu m$).}\label{MUX-S}	
    \end{figure}
    
	\begin{figure}
	\centering
	\includegraphics[width=3.4in,height=7.5in,keepaspectratio]{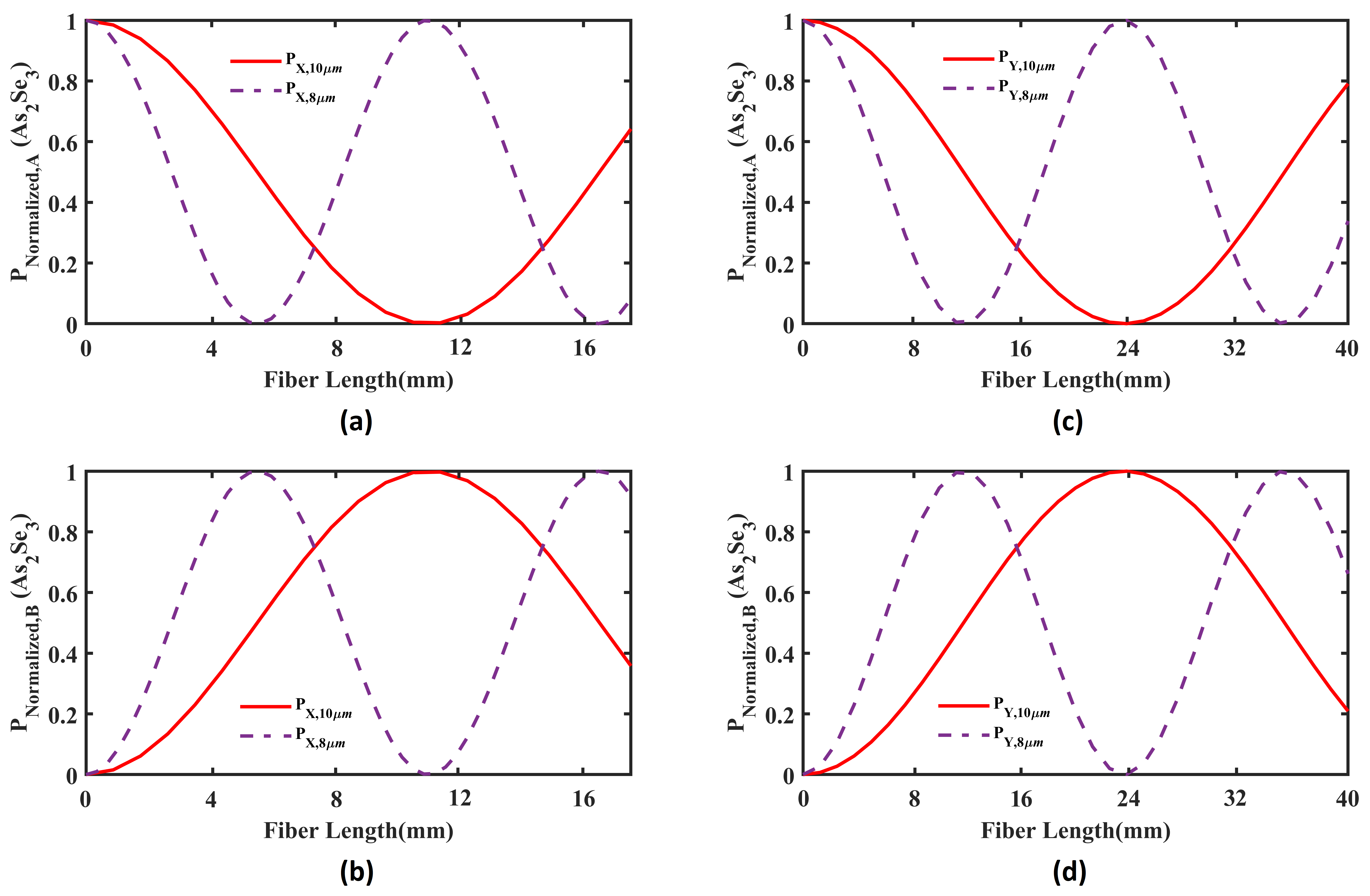} 
	\caption{Normalized output power of \ce{As2Se3} DC-PCF structure as a function of propagation length for: (a) X-mode, core $A$ ($\lambda = 8 \mu m$) (b) X-mode, core $B$ ($\lambda = 10 \mu m$) (c) Y-mode, core $A$ ($\lambda = 8 \mu m$) (d) Y-mode, core $B$ ($\lambda = 10 \mu m$).}\label{MUX-Se}	
    \end{figure}
    
     
	\begin{figure}
	\centering
	\includegraphics[width=2.5in,height=2.5in,keepaspectratio]{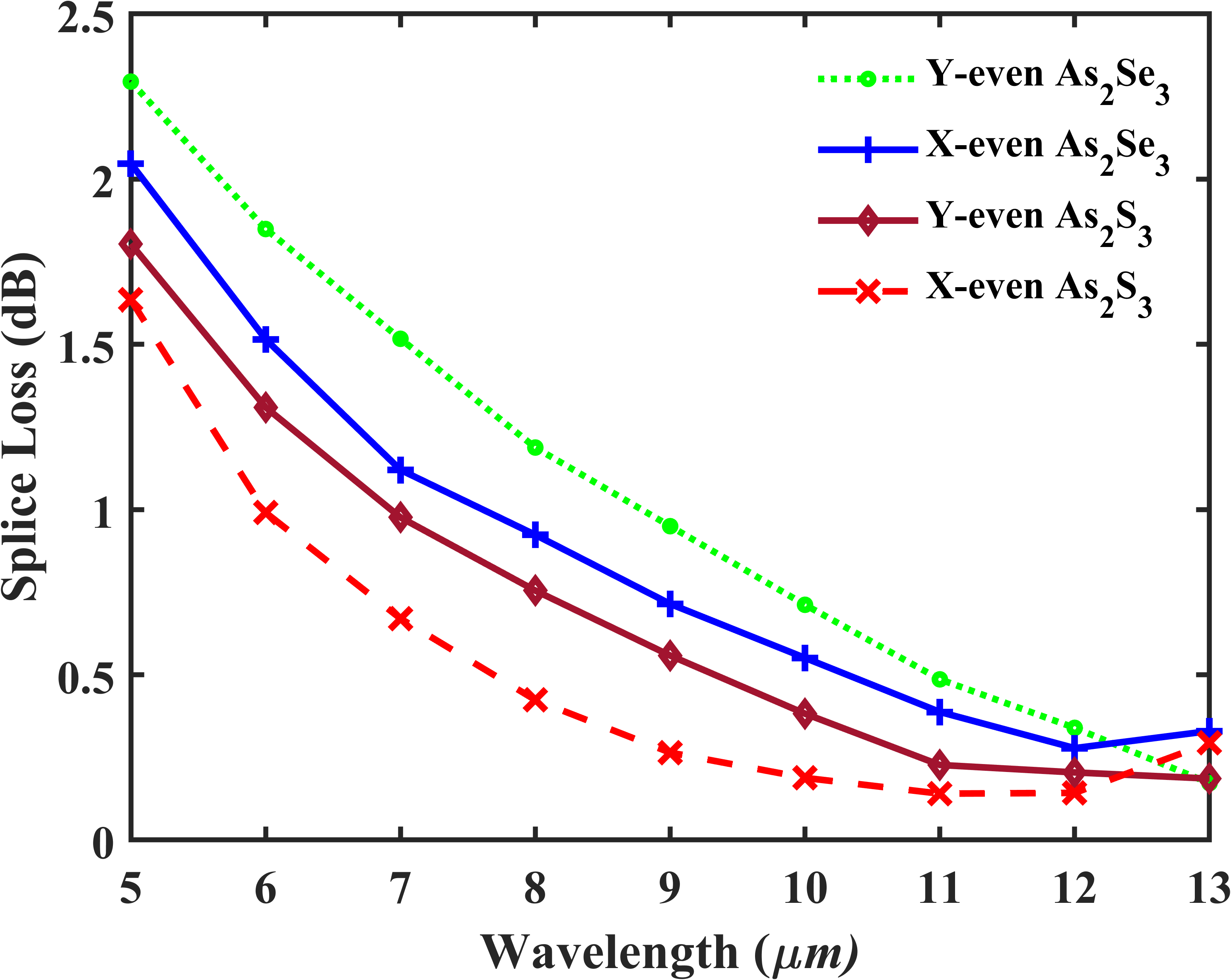} 
	\caption{Splice loss variation of DC-PCF structures with respect to wavelength for fundamental even supermodes.}\label{spl}	
    \end{figure}
    
	\begin{table*}
	\centering
	\caption{Comparison between \ce{As2S3} and \ce{As2Se3} DC-PCFs as multifunctional devices used in optical communications}\label{tab2}
\begin{tabular}{|c|c|c|}
			\hline
		
			\textbf{Applications/Parameter} & \textbf{\ce{As2S3 DC-PCF}} & \textbf{\ce{As2Se3 DC-PCF}}\\ \hline 
		     
			{\centering Polarization Splitter} & $\lambda$ = 9 $\mu m$ & $\lambda$ = 9 $\mu m$ \\
			     & Fiber length = 8.9 $mm$ & Fiber length = 14.8 $mm$ \\ \hline

{\centering WDM MUX-DeMUX} & separates $\lambda$ = 9 (X-Pol) \& 10 (Y-Pol) $\mu m$ & separates $\lambda$ = 8 (X-Pol) \& 9 (Y-Pol) $\mu m$ \\
			     & X-mode Fiber length = 3.99 $mm$ & X-mode Fiber length = 10.94 $mm$ \\
			     & Y-mode Fiber length = 8.28 $mm$ & Y-mode Fiber length = 23.96 $mm$ \\ \hline

			 {\centering Splice loss (X-even mode)} & 0.42 dB ($\lambda$ = 8 $\mu m$) & 0.92 dB ($\lambda$ = 8 $\mu m$) \\
			      & 0.26 dB ($\lambda$ = 9 $\mu m$) & 0.71 dB ($\lambda$ = 9 $\mu m$) \\ 
			     & 0.18 dB ($\lambda$ = 10 $\mu m$) & 0.55 dB ($\lambda$ = 10 $\mu m$) \\

			\hline
			
		\end{tabular}
	
\end{table*}

    \subsection{WDM MUX-DeMUX}
    
    Similar equations ((\ref{eq8}) and (\ref{eq9})) can also be used to evaluate the normalized output power for WDM MUX-DeMUX or wavelength splitters used in optical communications, and this normalized output power with the variation of wavelength for both chalcogenide materials are depicted in Fig. \ref{MUX-S} and Fig. \ref{MUX-Se}. The chalcogenide \ce{As2S3} DC-PCF separates 9 $\mu m$ and 10 $\mu m$ wavelengths of X-mode at a distance of $(25 \times 0.16 + 16 \times 0.249)/2 = 3.99$ $mm$ and of Y-mode at a distance of $(36 \times 0.23 + 23 \times 0.36)/2 = 8.28$ $mm$, see Fig. \ref{MUX-S}. On the other hand, from Fig. \ref{MUX-Se}, the \ce{As2Se3} DC-PCF separates 8 $\mu m$ and 9 $\mu m$ wavelengths of X-mode at a distance of $(11 \times 0.992 + 10 \times 1.1)/2 = 10.94$ $mm$ and of Y-mode at a distance of $(16 \times 1.496 + 15 \times 0.16)/2 = 23.96$ $mm$. Among the important findings is that DC-PCF is more effective as a WDM MUX-DeMUX in X-mode than in Y-mode due to the shorter fiber length.

    \subsection{Splice Loss}
    When two optical fibers are spliced together, some of its incoming optical power is not transmitted through the splice and is radiated out of the fiber, which is known as splice loss and calculated as \cite{8}:
     
     \begin{equation}
        P_{\text{splice Loss}} = -20\log_{10}\frac{2\omega^{}_{\text{SMF}} \omega^{}_{\text{DC-PCF}} }{ \omega^{2}_{\text{SMF}}  +  \omega^{2}_{\text{DC-PCF}}    } 
     \end{equation}
  
     From $ A_{\text{eff}} = \pi \omega^{2} $, we can find $\omega$, which is the mode field diameter of the fiber, and the mode field diameter of a single-mode fiber (SMF) is generally considered to be 10 $\mu m$.

     In Fig. \ref{spl}, we have shown the splice loss for each of the proposed elliptical DC-PCF structures for both of the chalcogenide materials. This figure demonstrates that \ce{As2S3} DC-PCF has lower splice loss compared to \ce{As2Se3} DC-PCF. Moreover, the splice loss (for X-even supermode) is around 0.42 dB, 0.26 dB, 0.18 dB for \ce{As2S3} and 0.92 dB, 0.71 dB, 0.55 dB for \ce{As2Se3} at the 8 $\mu m$, 9 $\mu m$, and 10 $\mu m$ wavelength windows, respectively. For practical optical communication applications, this is quite acceptable. Table \ref{tab2} is an overview of the findings for this study.

\section{\textbf{Conclusion}}\label{conclusion}
To summarize, we have designed a multifunctional hexagonal lattice DC-PCF structure incorporating elliptical air-holes for two types of chalcogenide materials (\ce{As2S3} and \ce{As2Se3}) showing highly birefringent and single-moded characteristics in the mid-IR (5-13 $\mu m$) wavelength windows, which is very important for optical communications. Results show that DC-PCF structures separate the two orthogonal modes, working as polarization splitters at 9 $\mu m$ wavelength, at a length of 8.9 $mm$ (for \ce{As2S3}) and 14.8 $mm$ (for \ce{As2Se3}). On top of that, \ce{As2S3} and \ce{As2Se3} DC-PCFs can also be utilized as WDM MUX-DeMUX for both X- and Y-polarization fundamental supermodes separating 9 $\mu m$/10 $\mu m$ and 8 $\mu m$/9 $\mu m$ wavelengths, respectively, where the X/Y-mode fiber lengths for \ce{As2S3} and \ce{As2Se3} are found 3.99 $mm$/8.28 $mm$ and 10.94 $mm$/23.96 $mm$, respectively. Moreover, the splice loss (for X-even supermode) is around 0.42 dB, 0.26 dB, 0.18 dB for \ce{As2S3} and 0.92 dB, 0.71 dB, 0.55 dB for \ce{As2Se3} at the 8 $\mu m$, 9 $\mu m$, and 10 $\mu m$ wavelength windows, respectively, which, for practical optical communication applications, is quite low and satisfactory. Hence, these chalcogenide DC-PCFs can be used in integrated and compact optical and photonic communications systems as short-length multifunctional devices. For the unique optical characteristics of these DC-PCFs, they can be utilized in mid-IR supercontinuum generation, surface plasmon resonance based PCF sensing applications, telecommunications, and frequency metrology.

\section{Acknowledgments}
No major funding was received for this research. Author declares no conflict of interest.


\begin{thebibliography}{00}

\bibitem{1}
H. Jiang, E. Wang, J. Zhang, L. Hu, Q. Mao, Q. Li, and K. Xie, Polarization splitter based on dual-core photonic crystal fiber, \href{https://doi.org/10.1364/OE.22.030461}{Opt. Exp. 22, 30461-30466 (2014)}.

\bibitem{2}
M. Hossain, B. Hossain, and Z. Amin, Small coupling length with a low confinement loss dual-core liquid infiltrated photonic crystal fiber coupler, \href{https://doi.org/10.1364/OSAC.1.000953}{OSA Cont. 1, 953-962 (2018)}.

\bibitem{3}
T. A. Birks, J. C. Knight, and P. Russell, Endlessly single-mode photonic crystal fiber, \href{https://doi.org/10.1364/OL.22.000961}{Opt. Lett. 22(13), 961–963 (1997)}. 

\bibitem{4}
M. Manimaraboopathy, G.A. Sathish Kumar, J. Mohanraj and M. Valliammai, Realization of all-optical multiplexer–demultiplexer in mid-IR wavelengths using triple-core photonic quasi-crystal fiber, \href{https://doi.org/10.1016/j.optcom.2020.126556}{Opt. Comm. 481, 126556, (2021)}. 

\bibitem{5}
L. Anbazhagan and G. Ganesan, Polarization-independent wavelength splitter based on silicon nanowire embedded dual-core pentagonal lattice photonic crystal fiber for optical communication systems, \href{https://doi.org/10.1117/1.OE.60.4.047103}{Opt. Eng. 60(4), 047103 (2021)}. 

\bibitem{6}
H. L. Chen, S. G. Li, Z. K. Fan, J. S. Li and Y. Han, A Novel Polarization Splitter Based on Dual-Core Photonic Crystal Fiber with a Liquid Crystal Modulation Core, \href{https://doi.org/10.1109/JPHOT.2014.2337874}{IEEE Photo. J. 6(4), 1-9, (2014)}.

\bibitem{7}
M. Rahman, A. Khaleque, T. Rahman and F. Rabbi, Gold-coated photonic crystal fiber based polarization filter for dual communication windows, \href{https://doi.org/10.1016/j.optcom.2020.125293}{Opt. Comm. 461, 125293, (2020)}.

\bibitem{choyon1}
A.K.M.S.J. Choyon and R. Chowdhury, Nonlinearity compensation and link margin analysis of 112-Gbps circular-polarization division multiplexed fiber optic communication system using a digital coherent receiver over 800-km SSMF link, \href{https://doi.org/10.1007/s12596-021-00714-x}{Journal of Optics 50, 512–521, (2021)}.

\bibitem{choyon2}
A.K.M.S.J. Choyon and R. Chowdhury, Design and performance analysis of spectral-efficient hybrid CPDM-CO-OFDM FSO communication system under diverse weather conditions, \href{https://doi.org/10.1515/joc-2021-0113}{Journal of Optical Communications, pp. 000010151520210113, (2021)}.

\bibitem{choyon3}
A. K. M. S. J. Choyon, S. M. R. Chowdhury and R. Chowdhury, Performance Analysis of a CPDM-QPSK Direct Detection Optical Transmission System under the effects of Cross-Polarization, \href{https://doi.org/10.1109/ICASERT.2019.8934494}{IEEE International Conference on Advances in Science, Engineering and Robotics Technology (ICASERT), Dhaka, pp. 1-4, (2019)}. 

\bibitem{choyon4}
A. K. M. S. J. Choyon, S. M. R. Chowdhury and S. P. Majumder, Performance of a CPDM QPSK Coherent Homodyne Optical Transmission System due to Cross Polarization Effects, \href{https://doi.org/10.1109/IC4ME2.2018.8465664}{IEEE International Conference on Computer, Communication, Chemical, Materials \& Electronic Engineering (IC4ME2), Rajshahi, pp. 1-5, (2018)}. 

\bibitem{choyon5}
A. K. M. S. J. Choyon, R. Chowdhury and S. M. R. Chowdhury, Performance Evaluation of CPDM 8-QAM for Faster Direct Detection Fiber Optic Communication System under the Effects of Cross-Polarization, \href{https://doi.org/10.1109/SPICSCON48833.2019.9065096}{IEEE International Conference on Signal Processing, Information, Communication and Systems (SPICSCON), Dhaka, pp. 1-4, (2019)}. 

\bibitem{choyon6}
A. K. M. S. J. Choyon, R. Chowdhury, S. M. R. Chowdhury and K. A. Taher, Cross-Polarization Induced Crosstalk Impact Analysis on the BER Performance of 100-Gbps CPDM 8-QAM CO-FOC System over Unrepeatered 100-km SSMF Link, \href{https://doi.org/10.1016/j.rio.2020.100012}{Results in Optics, 1, 100012, (2020)}.

\bibitem{choyon7}
A. K. M. S. J. Choyon and R. Chowdhury, Design of a 16×40 Gbps hybrid PDM-WDM FSO communication system and its performance comparison with the traditional model under diverse weather conditions of Bangladesh,  \href{https://doi.org/10.1515/joc-2020-0247}{Journal of Optical Communications, (2021)}.

\bibitem{choyon8}
R. Chowdhury and A. K. M. S. J. Choyon, Design of 320 Gbps Hybrid AMI-PDM-WDM FSO link and its Performance Comparison with Traditional models under Diverse Weather Conditions, \href{https://doi.org/10.1515/joc-2020-0135}{Journal Of Optical Communications, (2021)}.  

\bibitem{7a}
S. Chaudhary and A. Amphawan, Solid core PCF-based mode selector for MDM-Ro-FSO transmission systems, \href{https://doi.org/10.1007/s11107-018-0778-4}{Photon Netw Commun 36, 263–271, (2018)}.

\bibitem{8}
Z. Rahman, M. A. Rahman, M. A. Islam and M. S. Alam, Analysis of a Multifunctional Dual-Core Photonic Crystal Fiber for Optical Communications, \href{https://doi.org/10.1109/ICTP48844.2019.9041760}{2019 Int. Conf. Telecom. and Phot., IEEE, 1-4, (2019)}.

\bibitem{9}
Y. Zhao, S. Li, X. Wang, G. Wang, M. Shi, J. Wu. Design of a novel multi channel photonic crystal fiber polarization beam splitter, \href{https://doi.org/10.1016/j.optcom.2017.04.067}{Opt. Comm. 400, 79-83, (2017)}. 

\bibitem{10}
E. A. Hagras, M. F. O. Hameed, A. Heikal, and S. Obayya, Multifunctional photonic crystal fiber splitter for the two communication bands, \href{https://doi.org/10.1016/j.yofte.2019.101986}{Opt. F. Tech. 52, 101986, (2019)}.

\bibitem{11}
M. Sharma, N. Borogohain, and S. Konar, Index guiding photonic crystal fibers with large birefringence and walk-off, \href{https://doi.org/10.1109/JLT.2013.2281825}{J. Lightwave Technol. 31(21), 3339–3344, (2013)}.

\bibitem{12}
J. Lou, T. Cheng, and S. Li, Ultra-short polarization beam splitter with square lattice and gold film based on dual-core photonic crystal fiber, \href{https://doi.org/10.1016/j.ijleo.2018.10.109}{Optik, 179, 128–134, (2019)}.



\bibitem{13}
M. Yan, PL. Luo, K. Iwakuni, G. Millot, T. Hänsch and N. Picqué, Mid-infrared dual-comb spectroscopy with electro-optic modulators, \href{https://doi.org/10.1038/lsa.2017.76}{Light Sci Appl 6, e17076 (2017)}. 

\bibitem{14}
N. Picqué and T. Hänsch, Frequency comb spectroscopy, \href{https://doi.org/10.1038/s41566-018-0347-5}{Nat. Photon. 13, 146–157 (2019)}. 

\bibitem{15}
O. Benderov, I. Nechepurenko, B. Stepanov, T. Tebeneva, T. Kotereva, G. Snopatin, I. Skripachev, M. Spiridonov and A. Rodin, Broadband mid-IR chalcogenide fiber couplers, \href{https://doi.org/10.1364/AO.58.007222}{Appl. Opt. 58 (26), 7222–7226, (2019)}. 
	
	

\bibitem{16}
C. C. Huang and T.C. Sun, Numerical simulations of tunable ultrashort power splitters based on slotted multimode interference couplers, \href{https://doi.org/10.1038/s41598-019-49186-x}{Sci Rep 9, 12756 (2019)}. 


\bibitem{17}
F. Li, M. He, X. Zhang, M. Chang, Z. Wu, Z. Liu and H. Chen, Elliptical \ce{As2Se3} filled core ultra-high-nonlinearity and polarization-maintaining photonic crystal fiber with double hexagonal lattice cladding, \href{https://doi.org/10.1016/j.optmat.2018.03.025}{Optical Materials, 79, 137-146, (2018)}. 


\bibitem{18}
I. Chremmos, G. Kakarantzas and N. Uzunoglu, Modeling of a highly nonlinear chalcogenide dual-core photonic crystal fiber coupler, \href{https://doi.org/10.1016/j.optcom.2005.03.018}{Opt. Commun. 251 (4) 339–345, (2005)}. 


\bibitem{19}
Z. Fan, S. Li, J. Li, J. Wei and W. Tian, Ultra-bandwidth polarization splitter based on soft glass dual-core photonic crystal fber, \href{https://doi.org/10.1016/j.optmat.2015.04.052}{Opt Mater. 46, 384–8, (2015)}. 

\bibitem{20}
T. Zhao, S. Lou, X. Wang, M. Zhou and Z. Lian, Ultrabroadband polarization splitter based on three-core photonic crystal fiber with a modulation core, \href{https://doi.org/10.1364/AO.55.006428}{Appl. Optics, 55 (23), 6428–34, (2016)}. 



\bibitem{23}
C. Goncalves, M. Kang, B.-U. Sohn, G. Yin, J. Hu, D. Tan, and K. Richardson, New Candidate Multicomponent Chalcogenide Glasses for Supercontinuum Generation, \href{https://doi.org/10.3390/app8112082}{Appl. Sci. 8(11), 2082, (2018)}. 

 
\bibitem{24}
Z. Chen, L. Wan, J. Song, J. Pan, Y. Zhu, Z. Yang, W. Liu, J. Li, S. Gao, Y. Lin, B. Zhang, and Z. Li, Optical, mechanical and thermal characterizations of suspended chalcogenide glass microdisk membrane, \href{https://doi.org/10.1364/OE.27.015918}{Opt. Exp. 27, 15918-15925 (2019)}. 

\bibitem{25}
T. S. Saini, A. Kumar and R. K. Sinha, Broadband mid-IR supercontinuum generation in \ce{As2Se3} based chalcogenide photonic crystal fiber: A new design and analysis, \href{https://doi.org/10.1016/j.optcom.2015.02.049}{Opt. Comm. 347, 13-19, (2015)}. 

\bibitem{26}
H. Saghaei, M. Ebnali-Heidari, and M. K. Moravvej-Farshi, Midinfrared supercontinuum generation via \ce{As2Se3} chalcogenide photonic crystal fibers, \href{https://doi.org/10.1364/AO.54.002072}{Appl. Opti. 54(8), 2072-2079 (2015)}. 

\bibitem{27}
R. K. Sinha, A. Kumar and T. S. Saini, Analysis and Design of Single-Mode \ce{As2Se3}-Chalcogenide Photonic Crystal Fiber for Generation of Slow Light With Tunable Features, \href{https://doi.org/10.1109/JSTQE.2015.2477781}{IEEE J. Selected Topics in Quant. Elec. 22(2), 4900706, 287-292, (2016)}.


\bibitem{28}
T. S. Saini, A. Kumar and R. K. Sinha, Broadband Mid-Infrared Supercontinuum Spectra Spanning 2–15 $\mu m$ Using \ce{As2Se3}  Chalcogenide Glass Triangular-Core Graded-Index Photonic Crystal Fiber, \href{https://doi.org/10.1109/JLT.2015.2418993}{J. Lightwave Technol. 33(18), 3914-3920, (2015)}.

\bibitem{29}
A. Ben Salem, R. Cherif, and M. Zghal, Tapered \ce{As2S3} chalcogenide photonic crystal fiber for broadband mid-infrared supercontinuum generation, \href{https://doi.org/10.1364/FIO.2011.FMG6}{Front. in Opti. 2011/Laser Science XXVII, OSA Technical Digest, paper FMG6}. 



\bibitem{30}
TOPTICA Photonics, FemtoFiber dichro midIR, 2021. Available: \url{https://www.toptica.com/products/psfs-fiber-lasers/femtofiber-dichro/femtofiber-dichro-midir/}

\bibitem{31}	
M. J. Steel and R. M. Osgood, Elliptical-hole photonic crystal fibers, \href{https://doi.org/10.1364/OL.26.000229}{Opt. Lett. 26, 229-231 (2001)}. 	

\bibitem{32}	
K. Saitoh, Y. Sato, and M. Koshiba, Coupling characteristics of dual-core photonic crystal fiber couplers, \href{https://doi.org/10.1364/OE.11.003188}{Opt. Exp. 11, 3188-3195 (2003)}.

\bibitem{26a}
H. Saghaei, M. Ebnali-Heidari, and M. K. Moravvej-Farshi, Midinfrared supercontinuum generation via \ce{As2Se3} chalcogenide photonic crystal fibers, \href{https://doi.org/10.1364/AO.54.002072}{Appl. Opt. 54(8), 2072-2079 (2015)}. 

\bibitem{31a}
H. Balani, G. Singh, M. Tiwari, V. Janyani, and A. K. Ghunawat, Supercontinuum generation at 1.55 $\mu$m in \ce{As2S3} core photonic crystal fiber, \href{https://doi.org/10.1364/AO.57.003524}{Appl. Opt. 57, 3524-3533 (2018)}. 



\bibitem{33}

C. Dou, X. Jing, S. Li, J. Wu, and Q. Wang, A compact and low-loss polarization splitter based on dual-core photonic crystal fiber, \href{https://doi.org/10.1007/s11082-018-1516-y}{Opti. Quant. Elect. 50(6), 26, (2018)}.

\bibitem{34}
P.A. Agbemabiese and E.K. Akowuah, Numerical analysis of photonic crystal fiber of ultra-high birefringence and high nonlinearity, \href{https://doi.org/10.1038/s41598-020-77114-x}{Sci Rep 10, 21182 (2020)}. 

\bibitem{35}
H. Chen, G. Yan, E. Forsberg, and S. He, Terahertz polarization splitter based on a dual-elliptical-core polymer fiber, \href{https://doi.org/10.1364/AO.55.006236}{Appl. Opti. 55(23), 6236–6242, (2016)}.

\bibitem{36}
Z. Rahman, M. A. Rahman, M. A. Islam, and M. S. Alam, Design of an elliptical air-hole dual-core photonic crystal fiber for over two octaves spanning supercontinuum generation, \href{https://doi.org/10.1117/1.JNP.13.046013}{Journal of Nanophotonics, 13(4), 046013 (2019)}.













\end{thebibliography}
\end{document}